\documentclass[%
 reprint,
 amsmath,amssymb,
 aps,
 prl,
 longbibliography,
 lengthcheck,%
]{revtex4-1}

\usepackage{graphicx}
\usepackage{dcolumn}
\usepackage{bm}
\usepackage{hyperref}

\begin{document}


\title{Missing Transverse-Doppler Effect in Time-Dilation Experiments with High-Speed Ions}

\author{Santosh Devasia}
\affiliation{
 U. of Washington, Seattle, WA 98195-2600\\
}%


\begin{abstract}
Recent experiments with high-speed ions have investigated potential deviations from the time-dilation predicted by special relativity (SR). 
The main contribution of this article is to show that the SR predictions are matched by the experimental results only when the transverse Doppler effect in 
the observed emissions from the ions are neglected in the analysis. 
However, the Doppler effect in the emission cannot be neglected
because it is similar to the 
time dilation effect. Thus, the article 
highlights the need to consider Doppler emission effects  when validating SR time dilation using high-speed ion 
experiments. 
\begin{description}
\item[PACS numbers]
03.30.+p, 06.30.Ft, 42.62.Fi
\end{description}
\end{abstract}

\maketitle


\section{\label{sec:introduction}Introduction}
The interest in potential small Lorentz violations has led to recent interest in experimentally 
measuring the time-dilation predicted by special relativity (SR) using high-speed ions~\cite{Novtony_09}. 
It is noted that time-dilation predicted by special relativity (SR) has been verified  with a number of  
experiments starting with the classical experiment using Hydrogen canal rays by Ives and Stilwell~\cite{Ives_stilwell_1938}. 
More recent experiments have evaluated the time-dilation effect by using Doppler-shifted  lasers
to excite  transitions in high-speed Neon (Ne) and Lithium (Li) ions~\nocite{synder_hall_75,Kaivola_85,Mcgowan_93,Saathoff_03,Novtony_09}
\cite{Novtony_09,synder_hall_75}-\cite{Saathoff_03}.
A challenge in using high-speed ions is the presence of substantial Doppler broadening, which is resolved by using two lasers with different Doppler shifts that affect  ions at a unique speed. Emissions from the ions  are used to identify when  the
Doppler-shifted  laser frequencies  match the 
ion transitions, e.g., using the Lamb dip in the fluorescence spectrum~\cite{Saathoff_03} and to optimize the experiment~\cite{Kaivola_85,Mcgowan_93}. Then, the frequencies of the lasers used in the experiments  and the known transition frequencies of the ions are  used to validate SR predictions.  

\vspace{0.05in}
The main contribution of this article is to show that the 
SR predictions are matched by the experimental results only when the transverse Doppler effect in 
the observed emissions from the ions are neglected in the analysis. 
In particular, we review the Lamb-dip-based saturation spectroscopy experiments in~\cite{Saathoff_03} 
and show 
that the Doppler effect in the emission cannot be neglected
because it is similar to the 
time dilation effect. Thus, the article 
highlights the need to consider Doppler emission effects  when validating SR time dilation using 
high-speed ion experiments. 

\section{\label{sec:sat_spectroscopy}Saturation Spectroscopy with Lamb Dip}
\begin{figure}[!ht]
\begin{center}
\includegraphics[width=.6\columnwidth]{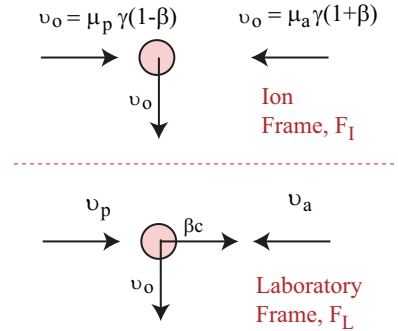}
\vspace{-0.1in}
\caption{SR without transverse Doppler effect in emissions. 
} 
\label{fig_SR_wo_doppler}
\end{center}
\end{figure}

A transition at frequency  ($\nu_o$)  in an ion 
moving with speed $v$
with respect to a laboratory frame $F_L$ 
can be excited by using 
parallel (co-propagating) or anti-parallel 
(counter-propagating) lasers as in Fig.~\ref{fig_SR_wo_doppler}. 
The associated  laser frequencies $\nu_p, \nu_a$ 
(parallel and anti-parallel in the laboratory frame) are given by the relativistic Doppler as~\cite{Saathoff_03}

\begin{eqnarray}
\nu_o = \nu_p \gamma (1 - \beta)  ~\approx \nu_p  (1 -\beta +\frac{ \beta^2}{2})
 \label{eq_mu_op}  \\
\nu_o  = \nu_a \gamma (1 + \beta) \approx \nu_a  (1 +\beta +\frac{ \beta^2}{2})
 \label{eq_mu_oa}    
\end{eqnarray} 
where the normalized speed $\beta = v/c$, the speed of light is $c$  and 
\begin{equation}
\gamma = 1/\sqrt{1 - \beta^2}.
\end{equation} 
The ratio 
$R$ , given by 
\begin{equation}
R = \frac{ \nu_p  \nu_a}{\nu_o^2} = 1 ,  \label{eq_mu_o_frac}
\end{equation}
is independent of  the speed of the ions. 
Potential dependence of the ratio  $R$  on speed $\beta$  is evaluated to test  SR.
The transitional frequency $\nu_o$ of the ions is known a-priori, and the frequencies $\nu_p, \nu_a $ 
are experimentally measured. 

\suppressfloats

The main challenge is to ensure that the lasers interact with ions that have the same speed $\beta$. For example, 
in  saturation spectroscopy~\cite{Saathoff_03},  one of the laser frequencies is kept constant and the other 
frequency is varied to observe the Lamb dip in the fluorescence spectrum at frequency $\nu_o$, which indicates that both lasers are acting on ions with the same speed.

\section{\label{sec:problem}The Problem: Doppler Effect in Emission}
\begin{figure}[!ht]
\begin{center}
\includegraphics[width=.6\columnwidth]{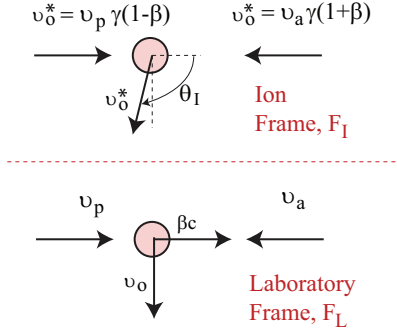}
\vspace{-0.1in}
\caption{SR with transverse Doppler effect in emissions. 
} 
\label{Fig_2_SR_w_Doppler}
\end{center}
\end{figure}

The problem is that previous analysis~\cite{Saathoff_03} does not consider the Doppler effect in the emissions observed from the ion. In particular, a photon  observed in the laboratory frame $F_L$  at a specific frequency   is 
emitted  from the ion at a different frequency with respect to a frame $F_I$ moving with the ion  as in Fig.~\ref{Fig_2_SR_w_Doppler}. 
\suppressfloats

\noindent
Since photodetectors 
detect emissions at frequency $\nu_o$ ~\cite{Saathoff_03}  perpendicular to the moving ions 
(in the laboratory frame $F_L$), the corresponding emission frequency$\nu_o^*$  in a frame $F_I$  moving with the 
ions is (from SR, e.g.,~\cite{Mansuripur_02}) 
\begin{equation}
\nu_{o}^* = \nu_{o} \gamma, 
\label{Eq_mu_o_mu_star}
\end{equation}
with angle $\cos{\theta_I} = -\beta$ in the ion's frame $F_I$. 
Even though counts in the  photodetectors (used to measure the emission) might be insensitive to 
the frequency of the photon, the observations will be sensitive to transverse Doppler effects if filters 
are used before the  photodetectors. For example, 
the emission is observed using  an interference filter (before the photodetector) centered at $\nu_o$ with  
a narrow  ($10$nm) halfwidth in~\cite{Reinhardt05} to precisely detect the laser frequency where the Lamb dip occurs --- this  makes the observations sensitive to the 
Doppler effect.

\vspace{0.05in}
The Doppler effect on emissions implies that the observed Lamb dip at 
frequency $\nu_o$ in the laboratory frame  $F_L$  corresponds to  
frequency $\nu_o^* $ in  frame $F_I$ moving with the ions.
Therefore, equations \ref{eq_mu_op}, \ref{eq_mu_oa} should be modified to 
\begin{eqnarray}
\nu_o^* = \nu_p \gamma (1 - \beta) \label{eq_mu_st_op} \\
\nu_o^*  = \nu_a \gamma (1 + \beta)  \label{eq_mu_st_oa}  
\end{eqnarray}
which yields a different expression for the ratio $R$ than Eq.~\ref{eq_mu_o_frac}
\begin{equation}
R ~= \frac{ \nu_p  \nu_a}{\nu_o^2} ~= \frac{ \nu_p  \nu_a \gamma^2}{(\nu_o^*)^2} ~= \gamma^2 
~= \frac{1}{1 -\beta^2} ~~ 
\approx 1 + \beta^2
 \label{eq_mu_o_st_frac}
\end{equation}

\vspace{0.1in}
Thus, if the Doppler effect is included in the emissions, then, from SR,  the ratio $R$  is 
no longer independent of  speed $\beta$. 
Current experimental results~\cite{Saathoff_03}  show that 
the ratio  $R$  is independent of speed $\beta$,  which 
would contradict SR prediction (in Eq.~\ref{eq_mu_o_st_frac})  if the emission Doppler effect is considered. 
Moreover, arguments (e.g., shift due to  electrical potentials in the ion beam) would be needed to clarify why the 
transition frequency shifts from $\nu_o$ to $\nu_o^*$ in the reference frame$F_I$  moving with the ion.

 \section{\label{sec:discussion}Doppler Emission Effect is not Negligible}
The Doppler effect on 
emission $\gamma$ (in Eq.~\ref{Eq_mu_o_mu_star})  is the same as the time dilation term 
$\gamma$ (in Eqs.~\ref{eq_mu_op}, \ref{eq_mu_oa})  that is being measured. 
Note that there are nonlinear second-order $\beta^2$  terms  in Eqs.~(\ref{eq_mu_op},\ref{eq_mu_oa}) if observations are at frequency $\nu_o$ in the frame $F_I$  moving with the  ions. With the addition of Doppler effect in emission (i.e.,  observations at frequency $\nu_o$ in the laboratory frame $F_L$), the expressions  in  Eqs.~(\ref{eq_mu_op},\ref{eq_mu_oa}) change to (by using Eqs.~(\ref{Eq_mu_o_mu_star}-\ref{eq_mu_st_oa})) 
\begin{eqnarray}
\nu_o = \nu_p  (1 - \beta) \label{eq_mu_lin_op} \\
\nu_o  = \nu_a (1 + \beta)  \label{eq_mu_lin_oa}  
\end{eqnarray}
that are  linear in the speed $\beta$  since the time-dilation effect $\gamma$ cancels out. 
Therefore, the emission Doppler effect 
is not negligible, and should be included in analysis of results from high-speed ion
experiments.  

\section{\label{sec:comparison}Effect of observation angle}
\begin{figure}[!ht]
\begin{center}
\includegraphics[width=.6\columnwidth]{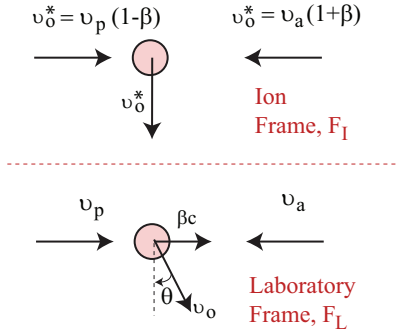}
\vspace{-0.1in}
\caption{Effect of observation angle.} 
\label{Fig_3_Obs_angle}
\end{center}
\end{figure}

Additional work is needed to resolve the apparent inconsistency of experimental observations of 
the ratio $R$ and  prediction from SR with Doppler effects in the emission. 
It is shown, below, that expressions for the  ratio $R$  
appear to be closer to the observed results if the angle of observation is not  
perpendicular to the motion of the ion. Note that previous work has shown that resonance fluorescence can be 
affected by the observational angle~\cite{Wang_Gao_97}. In the following analysis,  time dilation is not 
included, however, it is possible that similar arguments could be developed 
with the time-dilation effect. Additionally, predictions using such modifications of the observation angle 
would need to be validated experimentally. The linear Doppler 
shifts of the ion transition frequencies (without time dilation) are 
\begin{eqnarray}
\nu_o^* = \nu_p  (1 - \beta) \label{eq_mu_nSR_op} \\
\nu_o^*  = \nu_a (1 + \beta)  \label{eq_mu_nSR_oa}  
\end{eqnarray}
with the Doppler shift in the emission given by 
\begin{equation}
\nu_o = \nu_o^*\sqrt{1 + \beta^2}
\label{eq_mu_o_frac_nSR}
\end{equation}
where the emissions are observed at an angle $$\theta = \tan^{-1}(\beta)$$ from the perpendicular to the ion velocities as shown in Fig.~\ref{Fig_3_Obs_angle}, e.g.,~\cite{Devasia_light_09}. 
This leads to  nonlinear (second-order)  terms of speed  $\beta$ 
 in the Doppler shifted frequencies
\begin{eqnarray}
\nu_o = \nu_p  (1 - \beta)\sqrt{1 + \beta^2}  ~\approx  \nu_p (1 -\beta +\frac{ \beta^2}{2}) \label{eq_mu_nSR_op} \\
\nu_o  = \nu_a (1 + \beta)\sqrt{1 + \beta^2} ~\approx \nu_a (1 +\beta +\frac{ \beta^2}{2})  \label{eq_mu_nSR_oa}.  
\end{eqnarray}
that are similar to the second-order terms in Eqs.~(\ref{eq_mu_op},\ref{eq_mu_oa}), which 
will therefore match (upto second order in $\beta$) experimental results  for ion transitions~\cite{Kaivola_85,Mcgowan_93,Cyganski_07}. 

\suppressfloats

\noindent
The resulting ratio $R$ (using Eqs.~\ref{eq_mu_o_frac_nSR}-\ref{eq_mu_nSR_oa}) 
\begin{equation}
R ~= \frac{ \nu_p  \nu_a}{\nu_o^2} ~= \frac{ \nu_p  \nu_a }{(\nu_o^*)^2(1+\beta^2)} 
~= \frac{1}{1 -\beta^4} ~~ 
\approx 1 + \beta^4
 \label{eq_mu_o_nSR_frac}
\end{equation}
 is still not a constant ---  however,  it does not contain second-order terms of speed $\beta$  as in the results with SR and Doppler emission effect in  Eq.~(\ref{eq_mu_o_st_frac}). In this sense, the results with an observation angle $\theta$ from the perpendicular (as in Fig.~\ref{Fig_3_Obs_angle}) appear closer to the experimentally-observed, constant-ratio $R$ in Eq.~(\ref{eq_mu_o_frac}). Thus, it is possible, that the 
observation angle might be important in the interpretation of high-speed ion experiments.

\section{\label{sec:conclusions}Conclusions}
This  article  showed  that time-dilation predictions  of special relativity (SR) are matched by 
high-speed experimental results when the transverse Doppler effect in 
the observed emissions from the ions are neglected in the analysis. 
However,  the Doppler effect in the emission should not  be neglected
because it is similar to the 
time dilation effect. Thus, the article 
highlighted the need to consider Doppler emission effects  when using high-speed ion 
experiments to validate SR time-dilation predictions. 


%

\end{document}